# Reach and speed of judgment propagation in the laboratory

**Mehdi Moussaïd[a][1], Stefan M. Herzog[a], Juliane E. Kämmer[a,b], and Ralph Hertwig[a]**

[a]Center for Adaptive Rationality, Max Planck Institute for Human Development, 14195 Berlin, Germany

[b]AG Progress Test Medizin, Charité Medical School, 10115 Berlin, Germany

[1]Corresponding author: moussaid@mpib-berlin.mpg.de

**Keywords:** Collective behavior | Social influence | Social contagion | Diffusion chains | Reinforcement learning | Advice-taking

# Abstract

In recent years, a large body of research has demonstrated that judgments and behaviors can propagate from person to person. Phenomena as diverse as political mobilization, health practices, altruism, and emotional states exhibit similar dynamics of social contagion. The precise mechanisms of judgment propagation are not well understood, however, because it is difficult to control for confounding factors such as homophily or dynamic network structures. We introduce a novel experimental design that renders possible the stringent study of judgment propagation. In this design, experimental chains of individuals can revise their initial judgment in a visual perception task after observing a predecessor's judgment. The positioning of a very good performer at the top of a chain created a performance gap, which triggered waves of judgment propagation down the chain. We evaluated the dynamics of judgment propagation experimentally. Despite strong social influence within pairs of individuals, the reach of judgment propagation across a chain rarely exceeded a social distance of three to four degrees of separation. Furthermore, computer simulations showed that the speed of judgment propagation decayed exponentially with the social distance from the source. We show that information distortion and the overweighting of other people's errors are two individual-level mechanisms hindering judgment propagation at the scale of the chain. Our results contribute to the understanding of social contagion processes, and our experimental method offers numerous new opportunities to study judgment propagation in the laboratory.

# Significance statement

Individual judgments, feelings, and behaviors can spread from person to person in social networks, similarly to the propagation of infectious diseases. Despite major implications for many social phenomena, the underlying social contagion processes are poorly understood. We examined how participants' perceptual judgments spread from one person to another and across diffusion chains. We gauged the speed, reach, and scale of social contagion. Judgment propagation tended to slow down with increasing social distance from the source. Crucially, it vanished beyond a social horizon of three to four people. These results advance the understanding of some of the mechanisms underlying social contagion phenomena as well as their scope across domains as diverse as political mobilization, health practices, and emotions.

# Introduction

Social influence extends beyond people's direct contacts. In recent years, a large body of research has demonstrated that judgments, feelings, and behaviors can "spread" from person to person in social networks, similarly to the propagation of infectious diseases. Social contagion phenomena have been observed across a wide range of domains, such as public health (1–3), altruism and cooperation (4), risk perception (5), violence (6), political mobilization (7), and emotional states (8). Social contagion has been modeled in laboratory experiments (9) and described in field studies on subjects ranging from hunter–gatherer villages in Tanzania (10) to Facebook users (7, 8) and online recommendation networks (11). Behavioral contagion also serves adaptive functions in the collective behavior of animal swarms (12).

Unlike the propagation of diseases or the diffusion of information, the propagation of judgments often requires more than a single interaction between a sender and a receiver. Rather, the receiver needs to be "won over" by the sender's judgment before adopting and eventually spreading it further (13). Various social factors that reduce the receiver's uncertainty about the quality of the sender's judgment can thus facilitate judgment propagation (14). For instance, being exposed to the same judgment from different sources increases the probability of its adoption (13, 15, 16). Likewise, the sender's reputation shapes the strength of social influence (17), with reputation being treated as a proxy for judgment quality. Another factor is the history of past interactions between the sender and the receiver. Specifically, being able to repeatedly observe the good performance of a sender can reduce the receiver's uncertainty about the quality of the sender's judgment and consequently enhance the sender's social influence (18).

In a collective context, the repeated transmission of a judgment from person to person can give rise to global patterns of social contagion that can be described in terms of properties such as reach, speed, and intensity. In the context of innovation diffusion and viral marketing, simple mechanisms of influence can produce adoption cascades, in which the adoption of a product propagates across a large part of the population (11, 16, 19, 20). This form of social contagion, however, is restricted to environments in which people choose from a finite set of options and in which their choice is public and thus evident to all their neighbors. The spread of a *judgment* arises from a different and arguably more complex process, not least because judgments are often continuous and multidimensional. In addition, people might be only partially influenced by others' judgments (5). In fact, global cascades of judgment propagation are rarely observed. According to the three-degrees-of-influence hypothesis, judgment propagation is inherently limited to a social distance of about three people (21). That is, a person's influence on the judgment of neighboring individuals gradually dissipates with social distance and eventually ceases to have impact beyond three degrees of separation (1, 2, 4, 7). Several behavioral mechanisms have been proposed to explain this "wall", including the progressive deterioration of the judgment due to information distortion during social transmissions, and the fact that the social ties connecting people may not be robust across time, meaning that propagation pathways are interrupted (21). In addition, the nature of the transmission method—active teaching or simple copying, for instance—can foster or hinder propagation efficacy, as has been demonstrated in the domain of human cultural evolution (22, 23).

From the above literature, it appears that three main conditions should suffice for the emergence of long-distance judgment propagation: (i) The originator of the judgment consistently provides accurate judgments, so that observers have little uncertainty about the quality of his or her subsequent judgments; (ii) all individuals can *accurately* observe others' judgments and their quality, meaning that transmission noise is minimal; and (iii) the connections between people are stable, so that propagation pathways are uninterrupted. Under these conditions, an observer B should eventually be prompted to adopt the judgment of an initiator A, who repeatedly makes accurate judgments. In the absence of information distortion, B should in turn start performing well. This should subsequently prompt a third individual C to adopt B's—and thus indirectly A's—judgment. If the same process is repeated across many adjacent individuals in a chain, then the judgment of A could, in principle, propagate from person to person over long social distances. However, any systematic distortion of A's initial judgment over social transmission would make that judgment less accurate—and thus less persuasive. The increasingly distorted judgment would eventually cease to propagate. In this article, we examine whether the above three conditions are sufficient for the emergence of long-distance judgment propagation—and, if not, which other factors impede judgment propagation over long distances.

To this end, we present two studies. In Study 1, we investigated—using a simple visual perception task—the conditions under which the judgment of an individual A can spread to another individual B, as B repeatedly observes the judgment and performance of A. The conventional experimental paradigm in the advice-taking literature consists of observing one-shot interactions between two strangers—the participants neither know nor repeatedly interact with one another (14, 24). In contrast, we investigated how *repeated* interactions shape the strength of social influence, as has been done in work on "reputation formation" [see, e.g., (25, 26)]. In Study 2, we evaluated the collective dynamics of judgment propagation. We created unidirectional chains of six participants in which all individuals repeatedly interacted with their predecessor. That is, individual A initiated the chain, B repeatedly interacted with A, C repeatedly interacted with B, and so on up to individual F. Initially, all individuals were strangers. In each interaction round, they made an initial judgment, observed the judgment of their predecessor, potentially revised their judgment, and forwarded their final judgment to the next person in the chain. We then examined how far, under these favorable circumstances, A's judgment would travel down the chain. The results of Study 1 indicate that the influence of A on B increased gradually as the two got to know each other. However, as expected, this happened only if A performed considerably better than B. Study 2 showed that whenever individual A (i.e., the originator of the chain) consistently rendered more accurate judgments than the rest of the chain members, waves of judgment propagation were triggered down the chain: Round after round, A's judgments traveled increasingly farther. Crucially, however, the influence of A vanished beyond a social distance of about three people. That is, A influenced individuals B, C, and D, but not E and F. By modeling the experimental results and running numerical simulations, we then show that the participants' tendency to overweight their predecessor's error, combined with information distortion, generates *exponentially* increasing delays in propagation. This, in turn, resulted in the gradual extinction of influence with social distance.

## Results

**Experimental design**. We designed a visual perception task modeled on random-dot kinematograms (27). In each round, participants observed a set of 50 dots moving on a computer screen. Some dots moved consistently in a similar direction ("correlated dots") and others in random directions. Participants were asked to determine the main direction of the correlated dots as accurately as possible (see details in **Materials and Methods**, **Fig. 1,** and **Video S1**). The correct answer was a specific angle $\theta$ (i.e., a continuous value between 0° and 360°). Individuals were paired with the same partner for 15 consecutive rounds (see **Fig. 2**). In each round, individuals indicated their estimated angle by moving an arrow, then observed their partner's estimate, and finally either revised the first estimate or discounted the advice by pressing a button. We influenced the performance level by implementing three difficulty levels (i.e., adjusting the proportion of correlated dots; **Fig. 1**): low (level 1); medium (level 2); and high (level 3). In every round, the correct answer $\theta$ was always identical for both members of the pair, but the difficulty level could differ between them. For instance, one member of a pair might experience a low difficulty level (level 1) and the other a high difficulty level (level 3). In this example, the first individual would estimate the same true angle $\theta$ more accurately than the second. The difficulty level that each member of the pair experienced remained constant across the 15 rounds. Participants were not informed about the difficulty level they or their partner were facing. However, they could learn about their respective levels of performance by comparing their own and their partner's responses with the correct angle, which was displayed on screen at the end of each round. Consequently, participants could learn how accurate and thus how valuable their partner's judgment was.

**Individual behaviors.** In Study 1, we paired an individual facing a task of difficulty level *i* with a partner facing a task of difficulty level *j* for 15 consecutive rounds. In each round, we measured the partner's influence on the individual's judgments. We tested six conditions by varying *i* in {2,3} and *j* in {1,2,3}. Each *Si-Pj* condition (subject facing a difficulty level *i* paired with a partner facing a difficulty level *j*) was repeated with 100 participants.

First, we measured the deviation $\Delta_0^r = |x_0^r - x_p^r|$ and $\Delta_f^r = |x_f^r - x_p^r|$ between the individual's and the partner's estimate at round *r*, before and after social influence occurred. Here, $x_0^r$ corresponds to the individual's initial estimate at round $r$; $x_f^r$ to the individual's final estimate at round $r$ (after possibly revising the estimate); and $x_p^r$ to the partner's estimate at round $r$. For notational simplicity, we drop the superscript *r* when not referring to the number of a specific round. **Fig. 3A** shows the values of $\Delta_0$ and $\Delta_f$ at rounds 1 and 15 for all 100 participants in condition *S2-P1* (i.e., an individual facing a medium difficulty level paired with a partner facing a low difficulty level). In round 1, individuals tended to ignore the advice of their partner (i.e., $\Delta_f^1 \approx \Delta_0^1$), but by round 15 they were strongly influenced by it (i.e., $\Delta_f^{15} \approx 0, \forall \Delta_0^{15}$). We defined and studied three possible revision strategies (see the color coding in **Fig. 3A**): *ignore* (i.e., $\Delta_f = \Delta_0$), *adopt* (i.e., $\Delta_f$ is smaller than a small threshold value $d_a$), and *compromise* (i.e., $\Delta_f < \Delta_0$ and $\Delta_f \leq d_a$). We set the value of $d_a$ to 20°, which corresponded approximately to the length of the mouse pointer on the screen and thus tolerated minor inaccuracies in drawing an arrow. All data points for which $\Delta_0$ was lower than $d_a$ cannot support any conclusion and were thus not taken into account in this classification (but were included in all other analyses). **Fig. 3B** shows the observed proportion of participants adopting each strategy across the 15 rounds in all six conditions. When paired with a partner experiencing a low difficulty level (*S2-P1* and *S3-P1*), individuals switched rapidly from ignore to adopt within the first few rounds of interactions. This suggests that the initial strategy of disregarding a stranger's estimates was swiftly abandoned as an individual's uncertainty about the quality of the partner's performance diminished. The partner's influence increased faster for individuals facing tasks of high difficulty level 3 than for those facing a medium difficulty level 2 (see the comparison of the *S3-P1* and *S2-P1* adoption curves in **Fig. S1**). This suggests that individuals were sensitive to the *difference* between their own performance and their partner's performance, rather than just to their partner's absolute performance. Furthermore, individuals seemed to overweight their partner's error relative to their own (24, 25). For example, in condition *S3-P2*, individuals did not systematically adopt their partner's judgment, even though their partner performed better than they did. In this condition, the partner experienced a medium difficulty level 2 and thus often made substantial estimation errors. This impaired the individual's ability to detect that the partner's performance was better than their own. Similarly, individuals in condition *S3-P3* almost consistently ignored their partner's judgment, despite equal performance levels.

**Model**. Next, we modeled the above results using a computational model inspired by accounts of social reinforcement learning (28). The purpose was to establish a link between the mechanisms of influence within dyads observed in study 1 and the dynamics of judgment propagation that will be shown in study 2. For the model, we assumed that individuals assessed the performance or "quality" $Q^r$ of their partner at the end of each round *r*. Initially, in the absence of any information about their partner's performance, quality was set to the neutral value $Q^0 = 0$. At the end of each round *r*, individuals updated their perception of their partner based on the outcome of that round, as follows:

$$Q^r = Q^{r-1} + [\, e_0^r - \omega \cdot e^r\,],$$

where $e_0^r = |x_0^r - \theta^r|$ is the error of the initial estimate made by the individual at round *r*, and $e^r = |x^r - \theta^r|$ is the error made by the partner at round *r*. The partner's performance was therefore evaluated *relative* to the individual's own performance, as the results of study 1 suggested. The parameter $\omega$ is the error-overweighting coefficient, representing the weight that individuals assigned to their partner's errors. The model sets $\omega = 1$ if individuals weighted their own and their partner's errors equally, $\omega > 1$ if individuals overweighted their partner's errors relative to their own, and $\omega < 1$ if individuals underweighted their partner's errors relative to their own. Over time, individuals repeatedly updated the value of $Q^r$ and could form a positive ($Q^r > 0$) or negative ($Q^r < 0$) perception of their partner's quality, depending on the temporal sequence of relative errors experienced. After individuals observed their partner's judgment $x^r$ at round *r*, the current value of $Q^r$ determined which revision strategy they would select. The model stipulates two threshold values $\tau_1$ and $\tau_2$: It assumes that individuals ignore their partner's judgment if $Q^r < \tau_1$; adopt it if $Q^r > \tau_2$; and compromise if $\tau_1 \leq Q^r \leq \tau_2$. The final judgment is therefore $x_f^r = x_0^r$ if individuals ignore the advice; $x_f^r = (x_0^r + x^r)/2$ if they compromise; and $x_f^r = x^r + \delta$ if they adopt the advice. The value of $\delta$ accounts for the imperfection of the adoption strategy (as evident in **Fig. 3A**). We defined $|\delta|$ as a random value chosen in the interval [0 $d_a$], where $d_a$=20° (as in the strategy classification analysis); the sign of $\delta$ is negative if $x_0^r < x^r$ and positive if $x_0^r > x^r$. We determined the model parameters $\tau_1$, $\tau_2$, and $\omega$ by simultaneously fitting the model to the aggregate behavioral curves of all six conditions shown in **Fig. 3B**. The best fitting parameter values, minimizing the squared deviation from empirical data, were $\tau_1 = 15°$, $\tau_2 = 55°$, and $\omega = 2.0$. The fitted value of $\omega = 2.0$ is higher than 1, indicating a considerable overweighting of the partner's errors. The calibrated model predicts performance levels that are in good agreement with the data (**Figs. S2, S3, and S4**). For comparison, the same fitting procedure fixing $\omega = 1$ (i.e., assigning equal weight to the individual's and the partner's errors) failed to fully reproduce the observed performance levels (see **Figs. S3 and S4**).

**Judgment propagation**. In Study 2, we examined the dynamics of judgment propagation in chains involving six people, A, B, C, D, E, and F (see **Fig. 2B**). Each person in the chain was paired with their predecessor in the chain for 15 consecutive rounds. That is, in each round, the individual at position *k* observed the final estimate of the individual at position *k – 1* before entering her or his final estimate. This estimate, in turn, served as the advice given to the individual at position *k + 1*. The originator (individual A) had no partner and was thus never subject to social influence. A always experienced a low difficulty level 1, whereas all subsequent individuals always experienced a high difficulty level 3. At round 1, all individuals in the chain were strangers. However, we expected B to soon notice A's good level of performance and to adopt A's judgments. Consequently, we expected C to notice B's good level of performance and to indirectly adopt A's estimates. Round after round, we expected A's judgments to propagate increasingly further down the chain, with F eventually performing as well as A.

We collected data for 20 independent chains, each composed of six unique participants. In each round, we measured the influence of individual A on all subsequent individuals (B to F). Social influence *s* was measured as

$$s = 1 - (e_f - E_0^1)/(E_0^3 - E_0^1),$$

where $e_f = |x_f - \theta|$ is the observed error of the participant; the baseline error $E_0^i$ is the average error of all participants facing a task of difficulty level 1 before any interaction occurs. The social influence *s* thus measures the degree to which participants facing a high difficulty level 3 can minimize their error to that of the baseline error $E_0^1$ displayed by individual A (who faces a low difficulty level 1), scaled by $E_0^3 - E_0^1$. The values of *s* for all rounds and social distances, averaged across all 20 chains, are shown in **Fig. 4**. Waves of judgment propagation are clearly visible, indicating that propagation occurred. The judgment of A spread to B after two rounds, to C after four rounds, and even to D after about eight rounds. However, individuals E and F were rarely influenced by the judgment of A within the time horizon of the experiment (i.e., 15 rounds).

To better understand the mechanisms underlying this propagation pattern, we conducted simulations of the same experimental setup using our model (calibrated with the fitted parameter values from Study 1). We first evaluated the impact of each component of the model separately (**Fig. 5A**). As compared to a simple contagion model, the inclusion of information distortion curtailed the range of propagation to about 10 degrees of separation. This is obviously insufficient to explain the much shorter propagation ranges observed. However, *combining* information distortion with the overweighting of the partner's errors drastically limited the propagation range to around three to four degrees of separation, consistent with our data (see the analytical calculations in the supplementary information).

Finally, we used the model to explore the dynamics that take place over an extended time horizon. **Fig. 5B** shows the social influence of originator A over 100 rounds of interactions. The propagation range initially increased with the number of rounds but eventually plateaued at a social distance of approximately three degrees of separation for strong influence ($s > 0.7$) and four degrees of separation for moderate influence ($s > 0.4$). At five degrees of separation, social influence was no longer visible. We then calculated the time required until the originator's influence traveled different social distances. As **Fig. 5C** shows, the time required increased exponentially with social distance. That is, the time needed to reach the next individual in the chain was almost three times longer than the time needed to reach the previous one. The contagion wave thus *slowed down* with social distance, and traveling further than four individuals from the source thus required an excessively large number of rounds (see the propagation speed curves in **Fig. S5**). An analytical exploration of the model (presented in the supplementary information) revealed that each component of the model changed the functional form of the propagation delays. In the absence of information distortion and error overweighting, propagation time increased linearly with social distance. However, the inclusion of information distortion resulted in a polynomial increase in propagation times, and the addition of error overweighting resulted in the propagation time growing exponentially, consistent with our observations (**Figs. S6** and **S7**). Both of these factors impacted the functional form of the propagation delays.

## Discussion

Social contagion phenomena are difficult to study because they root in two intertwined layers of complexity, namely, that (1) induced by the network structure, and that (2) induced by the behavioral processes that operate within the network. In this article, we disentangled both layers and examined, in detail, the behavioral processes that drive judgment propagation for a simple network structure, namely, a transmission chain. Our experimental design enabled us to trigger waves of judgment propagation down the chain and study the behavioral factors influencing propagation.

For the time window of 15 interactions and a linear, unidirectional network structure examined here, we found that judgments rarely propagated beyond a distance of about three individuals. Our numerical and analytical results suggest that judgments could propagate farther but the time necessary to do so increases exponentially with social distance.

Propagation over distances greater than three to four individuals would require (1) a consistently more accurate originator, (2) an error-free observation of others' performance, and (3) a static chain structure to be maintained during several hundred of interactions.

Our results reveal some of the factors impeding spatial propagation *beyond* this range. First, judgments become progressively more distorted over successive social transmissions. Such distortion may even occur when individuals are able to accurately observe others' judgment and performance. In our behavioral data, distortion stemmed from imprecisions in imitating the partner's judgment or from slight deliberate alterations to the partner's judgment. Consequently, judgments tend to become gradually less accurate—and therefore less influential—as they travel further from their source. The second factor that impairs long-range propagation is the overweighting of other people's errors [also called "egocentric discounting" (25)]. This effect has previously been described for pairs in the advice-taking and reputation-formation literature (25); we have shown that it is also key to judgment propagation or lack thereof. In a pair, overweighting the partner's errors slows down the adoption of judgments because each error committed needs to be compensated for by several good performances. In a chain, the delay is amplified at each further position: The perceived quality of individual B in the eyes of individual C decreases as long as B performs poorly and thus as long as B has not yet adopted the judgment of A. For each subsequent step down the chain, the same process repeats with longer time delays. Overweighting other people's errors therefore contributes to the observed exponential growth of the time delays.

These results raise a number of interesting issues. Most importantly, future research should investigate how these behavioral dynamics may change in more complex network topologies (29). It is conceivable that multiple simultaneous sources of influence, as rendered possible in a more complex network, convey different judgments and may operate as sources of "noise". If so, such noise, in turn, is likely to impair propagation. Yet, clustered networks with redundant ties could also expose individuals to a converging judgment originating from different pathways and thus provide social reinforcement for its adoption (15). Our novel experimental design is versatile. It could be adapted to examine how judgments propagate through more complex network topologies and the extent to which presently observed limitedness of judgment propagation generalizes.

When combined with recent techniques assessing the structure of a social network (30), our approach could offer an experimental microworld that eventually can help to predict where and when the results of a promotion campaign will be most visible. Clusters of people separated by fewer than three individuals are likely to respond similarly to social interventions, giving rise to the clustering of judgments within regions of social networks (19). Additional simulations that we conducted show that good performers engender longer influence pathways (see **Fig. S8**). This "human factor" — that is, the initiator's past performance — influences the dynamics in addition to other structural factors related to the target's surrounding network structure, such as the node's degree or centrality (e.g. (15)). In sum, the complex interplay between human factors and the structure of the social network is key to better understand the dynamics of judgment propagation.

## Materials and Methods

**Experimental design**. The experimental task was to estimate the direction (i.e., angle) in which 50 dots, displayed in a circular area, moved on a computer screen (**Fig. 1**). The task was developed using HTML/Javascript and SmartFoxServer. A set of 25 unique angles equidistantly distributed between 0° and 360° served as true angles. The animation was displayed at a rate of 20 frames per second. While the dots were still moving, participants indicated their estimate by using the computer mouse to place a black arrow in the circular area. Once they had confirmed their first estimate (by pressing a mouse button), the movement of the dots was interrupted and a blue arrow indicating the partner's estimate was displayed. Participants could then use the mouse to revise the angle of their black arrow and pressed a button to confirm their final estimate. Participants who did not wish to revise their estimate could immediately press the confirmation button. Finally, the correct angle was displayed as a red arrow. Furthermore, the number of points $p$ scored by the participant in that round was displayed. The number of points awarded was an exponentially decreasing function of the error $p = 10exp(-5e_f^2)$ rounded to the closest integer, where $e_f$ is the error on the final estimate expressed in radians. Participants could therefore score between 0 and 10 points in each round. At the end of the experiment, the total number of points was converted into monetary bonuses at a rate of €0.005 per point.

**Procedure**. Data were collected between November 2015 and January 2016. The experimental procedure was approved by the Ethics Committee of the Max Planck Institute for Human Development. Fifteen participants were recruited for a preliminary study in which we collected the partners' estimates, and 100 participants were recruited for the main phase comprising studies 1 and 2 (58 males, 42 females, $M_{age}$ = 25.4 years; SD = 3.7). All participants gave informed consent. They received a flat fee of €10 for their participation, in addition to a monetary bonus depending on performance. In the preliminary phase, we collected independent estimates from 15 participants for each of the 25 true angles, and for each of the three difficulty levels (see **Fig. 1** for the implementation of the difficulty levels). Thus, each participant in the preliminary study provided one estimate for each true angle at each difficulty level (75 estimates in total). Partners' estimates in the main experiment (i.e., studies 1 and 2) were drawn from this pool of data.

In the main experiment, after receiving instructions, participants first completed five individual training rounds at each difficulty level. For study 1, they were then paired with a partner for 15 consecutive rounds (see **Fig. 2A**). The partner was chosen randomly among the participants in the preliminary study (see above). The 15 true angles and the corresponding estimates of the chosen partner were randomly drawn from the data provided by that partner during the preliminary phase. In each round, participants first estimated the angle, then observed their partner's estimate, and then indicated their final estimate. Subsequently, the correct angle was revealed, the participant's total score was updated, and a new round started. After 15 rounds, participants were informed that they would be paired with a new partner and another series of 15 rounds started. This procedure was repeated six times, corresponding to the six conditions *Si-Pj* with *i* in {2,3} and *j* in {1,2,3}, where *i* is the difficulty level faced by the participant and *j* is the difficulty level faced by the partner. As a manipulation check, we implemented a seventh condition *S1-P3*, in which the individual faced a low difficulty level and the partner faced a high difficulty level. As expected, participants in this condition systematically disregarded their partner's judgment because they could easily provide an accurate answer of their own. The results of this condition are not reported in the manuscript. For study 2, each participant participated in another series of 15 rounds in the chain setup (see **Fig. 2B**). A new chain was started every five participants (i.e., at participant numbers 1, 6, 11, and so on). In 15 consecutive rounds, the first participant in each chain (e.g., participant number 1) was paired with a partner (the originator) randomly drawn from the participants in the preliminary study. The 15 final estimates given by that participant served as the 15 pieces of advice that the next participant (participant 2) received. This procedure was repeated until the end of the chain was reached (e.g., until participant 5 was paired with participant 4) and a new chain started (e.g., participant 6 was paired with a random partner from the preliminary study). In this way, we collected a total of 20 independent chains. All participants in the chains experienced the highest difficulty level (except for the originator of the chain, who was drawn from the preliminary study and experienced a low difficulty level). In total, participants took part in eight series of 15 rounds: seven corresponding to the seven *Si-Pj* conditions in study 1, and one corresponding to the chain setup in study 2. The order of the eight conditions was randomized, that is, the position of each condition in the experiment was counterbalanced across participants. Finally, all participants completed eight additional series of 15 rounds for another study, in which they faced two or three partners at the same time. The results of this study are not reported in the present paper. The experiment lasted about 60 minutes in total.

# Acknowledgments

We thank Larissa Conradt and Winnie Poel for fruitful discussions and early contributions to the project. We are grateful to Anita Todd and Susannah Goss for editing the manuscript. This research was supported by a grant from the German Research Foundation as part of the priority program on *New Frameworks of Rationality* (SPP 1516) awarded to Ralph Hertwig (HE 2768/7-2). The funders had no role in study design, data collection and analysis, decision to publish, or preparation of the manuscript.

# Figures

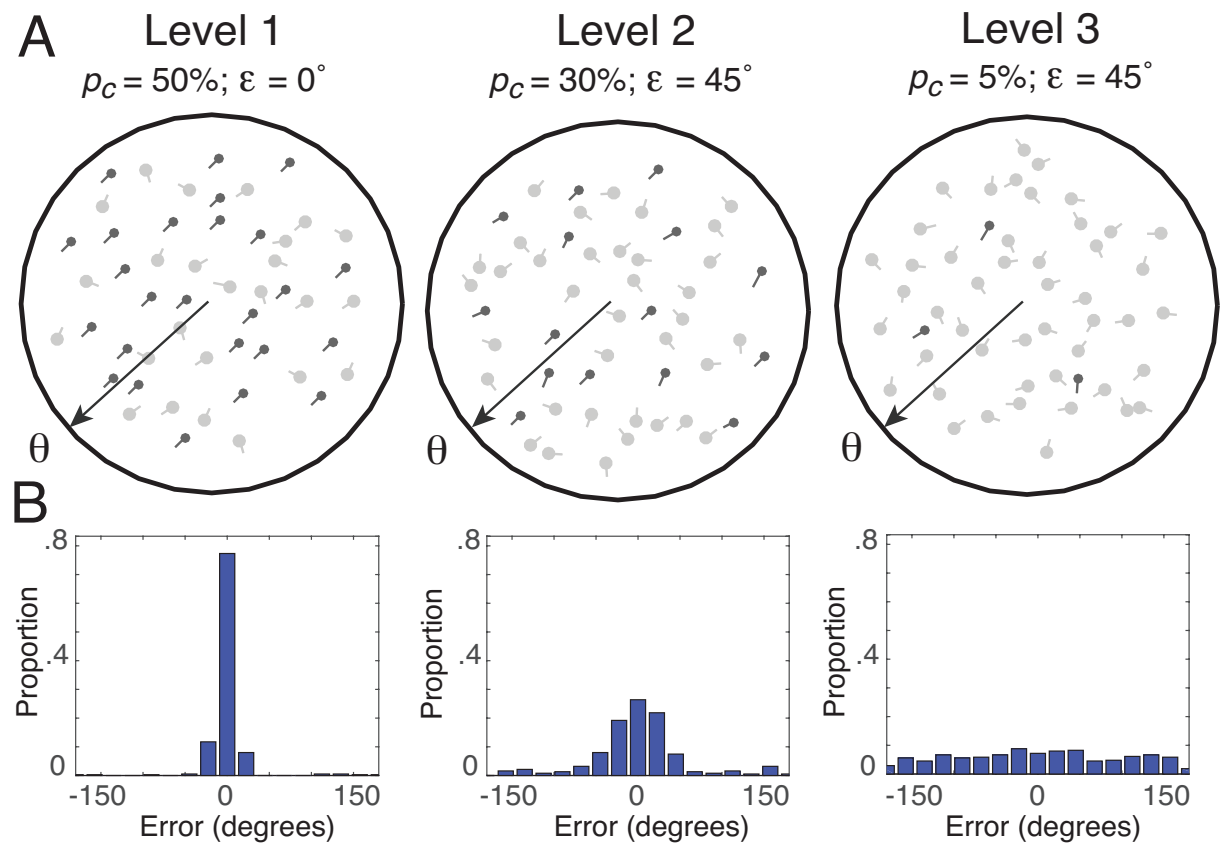

**Fig. 1. Experimental design.** (A) Participants were exposed to visual stimuli consisting of a set of 50 moving dots. A proportion $p_c$ of correlated dots moved in the direction $\theta + \bar{\varepsilon}$, and the remaining proportion $1-p_c$ of uncorrelated dots moved in random directions. Here, $\theta$ was the true value that participants had to estimate visually and $\bar{\varepsilon}$ was a random deviation from the true value, sampled from the interval $[-\varepsilon\ \varepsilon]$ for each correlated dot (to avoid exactly parallel trajectories of the correlated dots, which would have rendered the task too easy). The three difficulty levels were characterized by different proportions of correlated dots, $p_c$ = 50%, 30%, and 5%. In addition, the deviation was set to $\varepsilon = 0°$ in difficulty level 1 and to $\varepsilon = 45°$ in difficulty levels 2 and 3. The black arrows show the true value $\theta$ in each example and were not visible to participants. During the experiment, correlated and uncorrelated dots appeared in the same color (light and dark grey are used in this figure for illustration only). See **Video S1** for animated examples. (B) The distribution of errors made by all 100 participants at each difficulty level in the absence of social influence (i.e., before any interaction occurred). The values of $p_c$ and $\varepsilon$ for each difficulty level were calibrated prior to the experiment to yield clearly distinguishable error distributions.

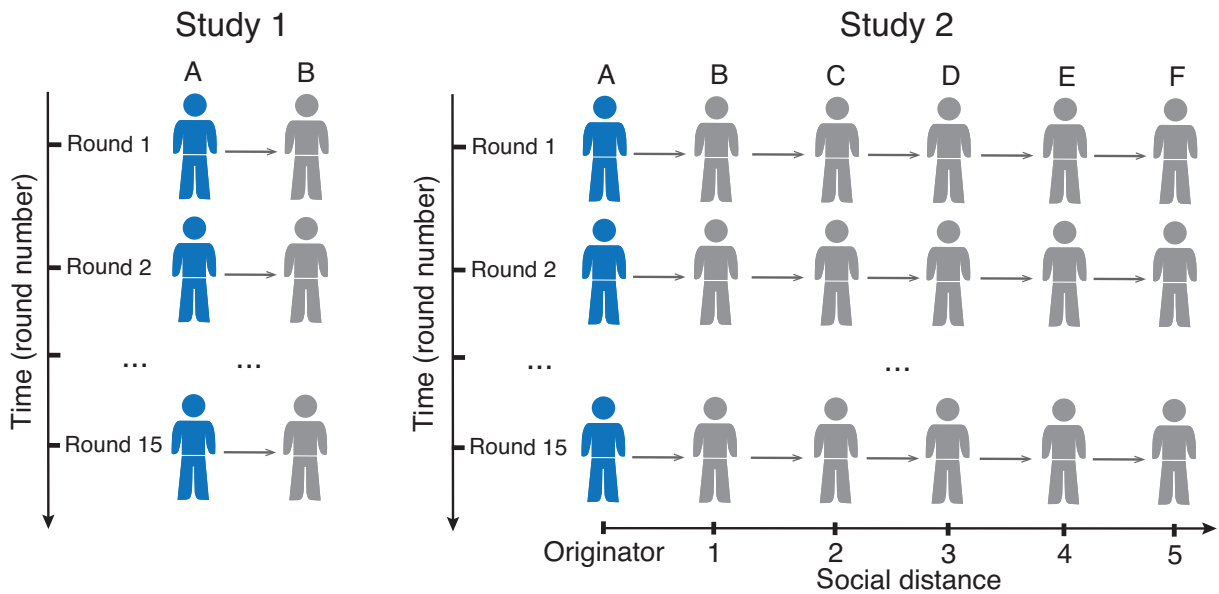

**Fig. 2. Schematic representations of the two experimental setups.** In study 1, two individuals A and B were paired for 15 consecutive rounds. In each round, they separately faced the visual perception task described in **Fig. 1**. The correct answer $\theta$ was always identical for both of them, but the difficulty level could differ between them. The respective difficulty levels were fixed across the 15 rounds. In each round, individual B could reconsider her or his initial estimate after observing A's estimate. (B) In study 2, six participants (here called individuals A, B, C, D, E, and F) were connected along a chain for 15 consecutive rounds. Each individual in the chain was paired with his or her predecessor in the chain. Otherwise, the same procedure was followed as in study 1. That is, in each round, the estimate of the originator, individual A (in blue), was shown to individual B, who could then reconsider her or his initial estimate accordingly. Individual B's revised estimate was then shown to C, who could, in turn, reconsider her or his estimate, and so on until individual F. This process of social transmission from A to F was repeated across the 15 rounds. The originator A always experienced a low difficulty level 1 and all subsequent participants always experienced a high difficulty level 3.

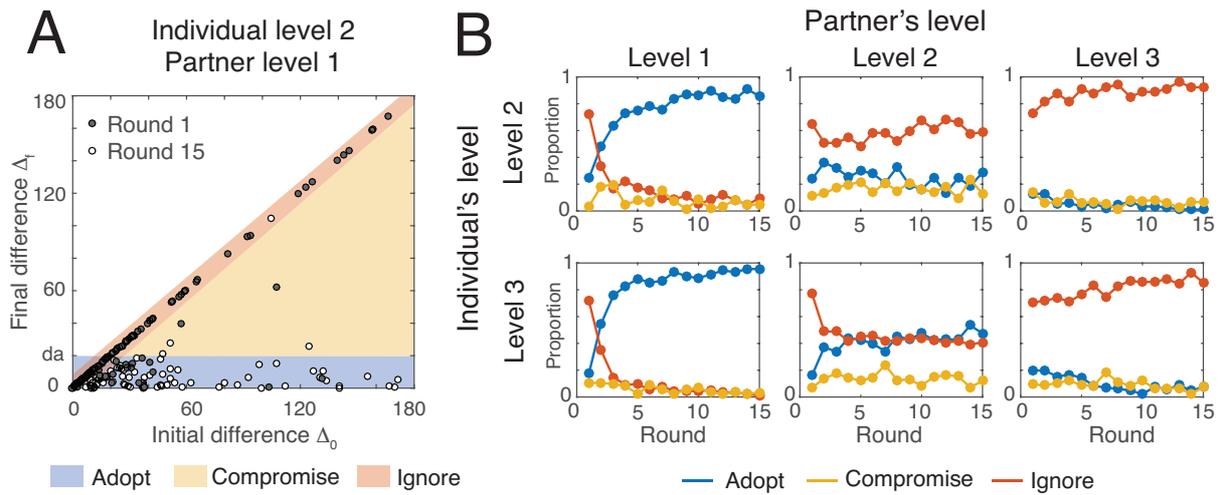

**Fig. 3**. **Participant behavior in study 1**. (A) Illustration of the behavioral change between round 1 (black dots) and round 15 (white dots) for all 100 participants in condition *S2-P1* (individuals, *S*, facing a medium difficulty level 2 paired with partners, *P*, facing a low difficulty level 1). The three background colors indicate the zones corresponding to the three strategies: ignore (red), compromise (yellow), and adopt (blue). Whereas participants mostly ignored their partner's judgment at round 1 (72% of black points in the red zone), they tended to adopt it at round 15 (85% of white points in the blue zone). (B) The proportion of participants adopting each strategy across the 15 rounds in all experimental conditions. The detailed example shown in A corresponds to the first and last rounds in the upper left subpanel of B.

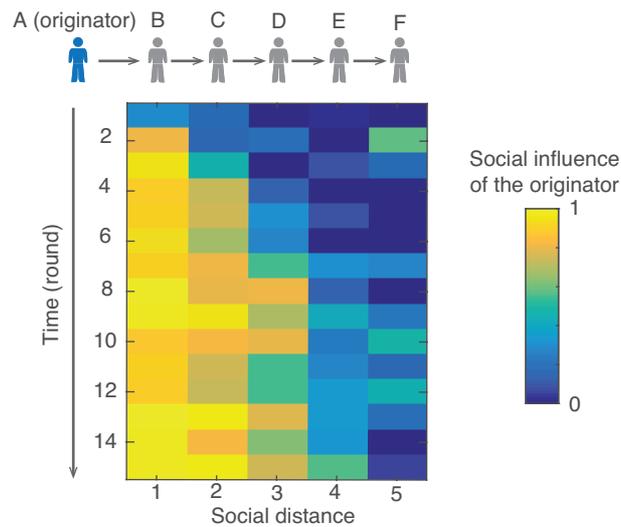

**Fig. 4. Judgment propagation down the chain (study 2)**. Observed intensity of social influence averaged over 20 experimental chains. The color coding indicates the influence of the originator's judgment on the final estimates of all other individuals in the chain, as a function of their social distance from the originator (*x* axis) and the number of the round (*y* axis; see also **Fig. 2B**). Individuals located one degree of separation from the originator (i.e., social distance = 1) rapidly adopted the originator's judgments (social influence approached 1 as early as round 2). Individuals located two degrees of separation from the originator were influenced by the originator after about 4 rounds of interaction. Participants located at a social distance greater than 3 were rarely influenced by the originator.
(See high quality figure in the final publication)

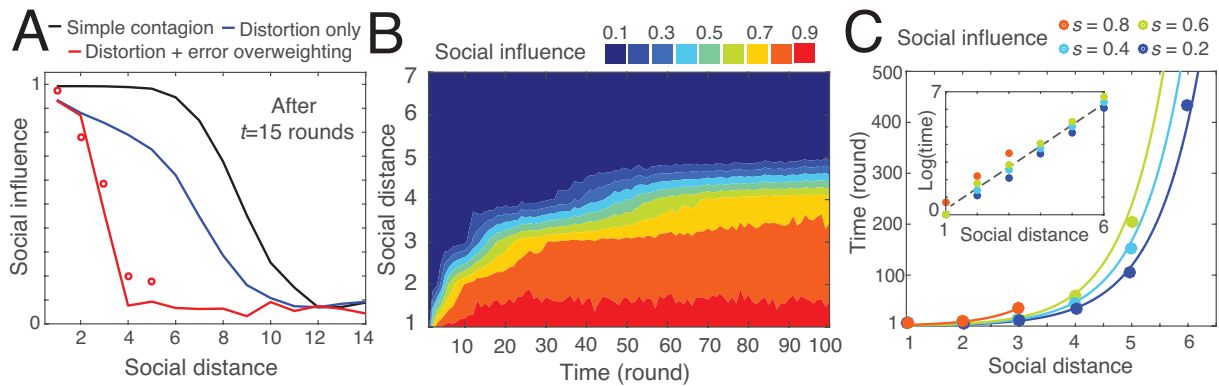

**Fig. 5**. **Dynamic features of judgment propagation in study 2**. (A) The intensity and range of social contagion after 15 rounds, assuming different propagation mechanisms (model previously calibrated in study 1): full and immediate adoption of the partner's judgment (in black), distortion of information only (in blue), distortion of information and error overweighting (in red). Circles correspond to experimental data. (B) Intensity and range of social contagion over 100 rounds of interactions as predicted by the model. The range of social contagion initially increased with time but eventually plateaued at a social distance of 4. (C) The time delay required for social influence to travel through the chain increased exponentially with social distance. The dots indicate the model predictions (for 1,000 simulation runs), and the lines show the best exponential fit. For distances larger than 4 persons, the time delay exceeded 100 rounds before a weak influence ($s = 0.2$, in blue) appeared. Strong influence ($s = 0.8$, in red) could not propagate beyond a social distance of 3 persons because information distortion rendered the (indirect) adoption of the originator's judgment increasingly impossible. The inset shows the same figure with a logarithmic transformation along the *y* axis, confirming the exponential growth of the time delay.